\documentclass[prl,twocolumn,showpacs,superscriptaddress]{revtex4-1}

\usepackage{graphicx}
\usepackage{dcolumn}
\usepackage{amsmath}
\usepackage{epsfig}
\usepackage{bm}
\usepackage{amssymb}
\usepackage{float}
\usepackage{natbib}
\usepackage{color}
\usepackage{hyperref}
\hypersetup{colorlinks=true,linkcolor=blue,citecolor=blue,urlcolor=black}

\begin{document}

\title{Microwave probes Dipole Blockade and van der Waals Forces in a Cold Rydberg Gas}

\author{R. Celistrino Teixeira}
\author{C. Hermann-Avigliano}
\author{T.L. Nguyen}
\author{T. Cantat-Moltrecht}
\affiliation{Laboratoire Kastler Brossel, Coll\`ege de France, ENS, UPMC Univ Paris 06, CNRS,\\ 11 place Marcelin Berthelot, 75005, Paris, France.}
\author{J.M. Raimond}
\author{S. Haroche}
\author{S. Gleyzes}
\author{M. Brune}
\affiliation{Laboratoire Kastler Brossel, Coll\`ege de France, ENS, UPMC Univ Paris 06, CNRS,\\ 11 place Marcelin Berthelot, 75005, Paris, France.}

\email{michel.brune@lkb.ens.fr}

\date{\today}

\pacs{03.67.-a,32.80.Ee,32.30.-r}


\begin{abstract}
We show that microwave spectroscopy of a dense Rydberg gas trapped on a superconducting atom chip in the dipole blockade regime reveals directly the dipole-dipole many-body interaction energy spectrum. We use this method to investigate the expansion of the Rydberg cloud under the effect of repulsive van der Waals forces and the breakdown of the frozen gas approximation. This study opens a promising route for quantum simulation of many-body systems and quantum information transport in chains of strongly interacting Rydberg atoms.  
\end{abstract}

\maketitle

The strong dipole-diplole interaction between cold Rydberg atoms is the focus of an intense theoretical and experimental interest~\cite{MX_GALLAGHERREVIEWRYD08,MX_SAFFMANRYDREVIEW10,MX_PFAURYDBERGTUTO12}. It provides an efficient platform for quantum information processing, with quantum gates based on the manipulation of single trapped atoms~\cite{MX_GRANGIERRYDENTANGLE10,MX_SAFFMANCNOT10}. It leads to optical non-linearities~\cite{MX_ADAMSCOLLRYDBLOCKADE11,MX_GRANGIERNONLINEARRYD12}, even at the single photon level ~\cite{MX_VULETICPHOTONSRYD13,MX_ADAMSRYDBSTORAGE13,MX_HOFFERBERTHRYDTRANS14,MX_REMPERYDTRANS14}. Finally, it opens the way to quantum simulators of self-organization processes and phase transitions~\cite{MX_LUKINBLOCKADECRYSTAL10, MX_LEEANTIFERRO11,MX_ZOLLERSIMULREVIEW12, MX_ZOLLERRYDBERGSIMUL12, 
MX_FLEISCHHAUERANTIFERRO14}.

A variety of methods have been implemented to investigate the dipole-dipole interaction. An early observation of a Rydberg excitation line broadening at high density provided a direct evidence of the interaction~\cite{ENS_DENSEGAS}. Dipole blockade~\cite{QI_LUKINDIPOLEBLOCKADE01,MX_PFAURYDCOLL07,MX_SAFFMANBLOCKADE09,MX_GRANGIERBLOCK09} was demonstrated with individually trapped atoms~\cite{MX_BROWAYESBLOCKADE14}, observed on Rydberg atom counting statistics in atomic ensembles~\cite{MX_PILLETCOUNTSTAT12,MX_WEIDEMULLERRYDAGREGATES13,MX_SAFFMANATOMICFOCK14,MX_OTTSUPERATOM15} or on collective single-photon emission~\cite{MX_KUZMICHFULLBLOCKADE12}. Spatial self-organization has been evidenced by individual atom imaging in gases~\cite{MX_RAITHELRYDBERGIMAGE11} and in optical lattices~\cite{MX_BLOCHSTRUCTUREDRYDBERG12}.

We present here a direct measurement, based on microwave spectroscopy, of the interaction energy distribution in a cloud of ultra-cold Rydberg atoms in the blockade regime. We use it to observe the atomic cloud expansion driven by the repulsive van der Waals (vdW) forces, revealing the limits of the frozen gas approximation~\cite{MX_PILLETCOLDRY98}. The observations are in good agreement with Monte Carlo simulations.

\begin{figure}[tbm]
\centering
\includegraphics[width=0.4\textwidth]{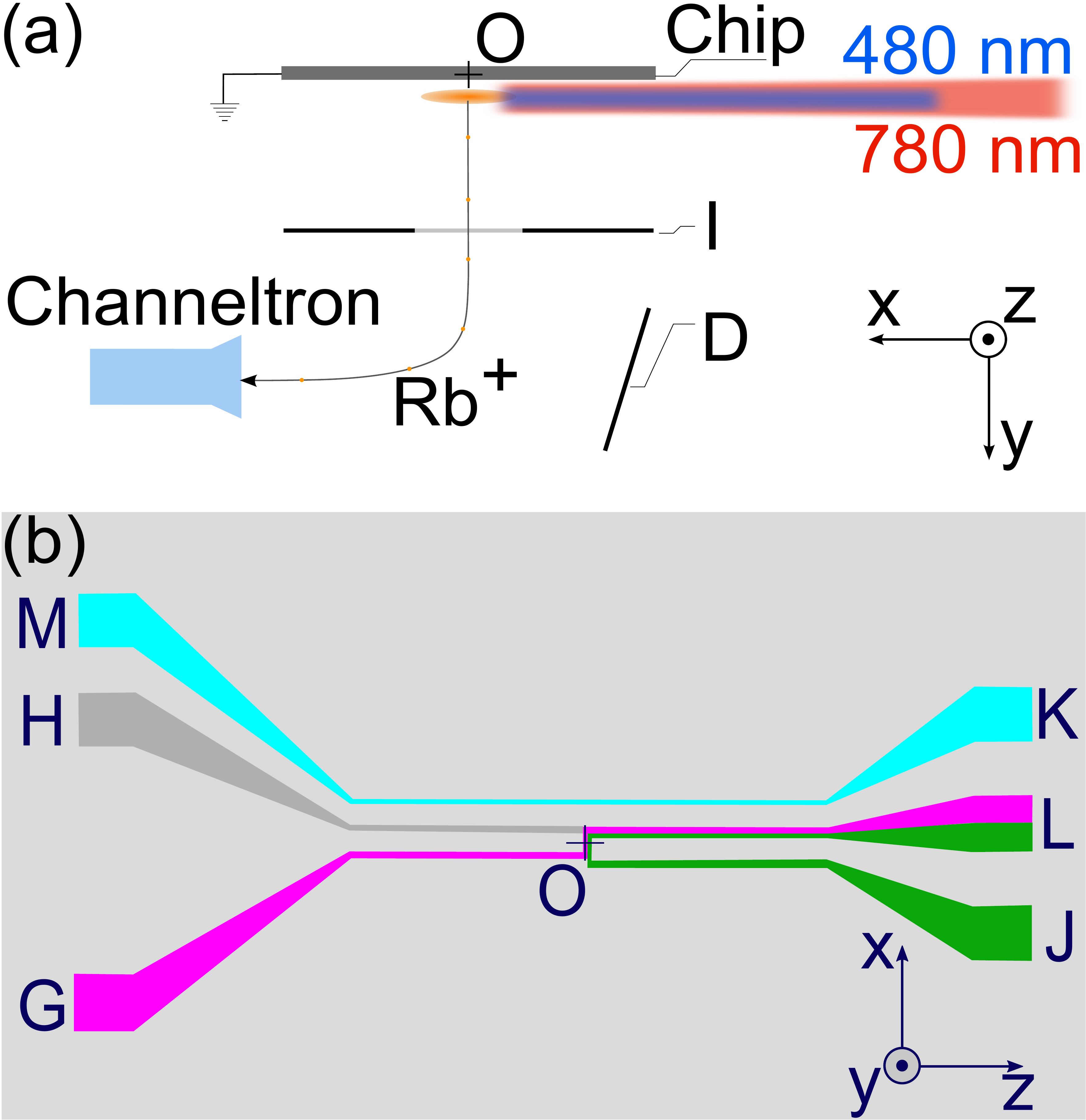}
\caption{(color online) (a)  Scheme of the central part of the setup. (b) Scheme of the superconducting atom chip. The axes are defined in the two panels. The origin $O$ is at the center of the $Z$ wire $LG$.} 
\label{scheme}
\end{figure}

The core of the set-up is sketched on figure \ref{scheme}(a) (details in~\cite{ENS_BECCHIP08,ENS_CHIPSPECTRO14}). Rubidium atoms are trapped on a superconducting atom chip, which is cooled down to $4.2\,$K. In the experimental sequence, they first effuse from a 2D-Magneto Optical Trap (MOT) placed at room temperature outside the cryostat. They are then caught a few millimeters away from the reflecting front chip surface in a mirror-MOT at a temperature $ T\approx 100\ \mu$K. The MOT quadrupolar magnetic field is generated by centimeter-sized superconducting coils. The atoms are then transferred into a compressed mirror-MOT located 700~$\mu$m away from the chip. Its magnetic field is the superposition of that produced by the $U$-shaped superconducting wire connecting pads $J$ and $L$ [Fig. \ref{scheme}(b)] with a uniform bias. 

The magnetic field is then transiently switched off while the atoms are cooled down to $12\ \mu$K by an optical molasses. They are optically pumped into the $5S, F=2,m_F=2$ level and transferred into a Ioffe-Pritchard trap. Its field is a superposition of that generated by the $Z$-shaped $GL$ wire (width: $70\ \mu$m) with a uniform bias. The field at the bottom of the trap is aligned along the $x$ quantization axis (defined in Fig. \ref{scheme}). The atoms are then evaporatively cooled just above the BEC transition using a radio-frequency field radiated by the $KM$ wire~\cite{ENS_BECCHIP08}. The total cloud preparation time is about 8 seconds. The final thermal cloud, $300\ \mu$m away from the chip surface, contains $\approx 12\,000$ atoms at a temperature of about $500\,$nK. The Gaussian atomic density peaks at $2.4\ 10^{12}\ $cm$^{-3}$, with dispersions $2\sigma_x=44\ \mu$m and $2\sigma_y \simeq 2\sigma_z = 11\ \mu$m  along the three axes. This moderate density makes the interaction between the Rydberg atoms and the ground state cloud (a few tens of kHz~\cite{MX_PFAURYDBERGBEC13}) negligible as compared to the vdW interaction between Rydberg atoms. 

A large fraction of the atoms released from the trap is adsorbed on the gold front chip surface. The deposited atoms  create large stray electric field gradients in the trap. This  detrimental effect is avoided by coating the full chip surface with a metallic rubidium layer, stable at 4~K, which increases dramatically the Rydberg levels coherence time~\cite{ENS_CHIPSPECTRO14}.
 
We induce the two-photon $5S\rightarrow 60S$ transition by 2~$\mu$s 780 nm `red' and 480~nm `blue' laser pulses. The detuning with respect to the intermediate $5P_{3/2}$ level is 540~MHz. The red and blue lasers propagate along the $x$ axis [Fig. \ref{scheme}(a)], with $150$ and $22\ \mu$m waists, 50~$\mu$W and 8~mW powers, and $\sigma_+$ and $\sigma_-$ polarizations  respectively. They excite the $60S_{1/2},m_j=1/2$ sublevel with a $400$~kHz two-photon Rabi frequency. Ten excitation pulses, at a 3~ms time interval, are sent on the atomic cloud without noticeable heating. 
 Both lasers are frequency-locked to a  transfer cavity, whose length is stabilized to a rubidium saturated absorption line.  At low Rydberg density, the excitation linewidth  is $\gamma=600$~ kHz, after minimization of the residual electric field in the $y$ direction using electrode $I$ facing the chip (kept at 0~V) [Fig. \ref{scheme}(a)]. The contribution of the residual electric field gradients is below 50~kHz~\cite{ENS_CHIPSPECTRO14}. 

We detect the Rydberg atoms by field-ionization [Fig. \ref{scheme}(a)], with a 90$\pm$10\% detection efficiency~\cite{ENS_CHIPSPECTRO14}. An electric field ramp, produced by electrode $I$, reaches at different times the ionization thresholds of the 60S and 57S levels involved in the spectroscopy (37 V/cm and 41~V/cm respectively). The resulting ions are accelerated and deflected (electrode $D$) towards a channeltron counter (Sjuts Optotechnik KBL 10RS-EDR). For short delays between laser excitation and detection, the ion signal is broadened, an effect we attribute to state mixing in the field ramp in the presence of dipole-dipole interactions. We thus apply the ionizing field 150$\mu$s after the laser pulse, leaving time for the Rydberg cloud to expand due to the repulsive vdW forces (see below). Atomic interactions are thus negligible during field ionization. A residual partial overlap of the ionization signals of 60S and 57S results in a transfer rate offset, which is measured and subtracted from the data. 

For interatomic distances, $R$, larger than $\simeq 3\ \mu$m, the dipole-dipole energy shift for two 60S atoms reads $hC_{60,60}/R^6$. A numerical diagonalization of the Hamiltonian provides a repulsive vdW interaction with $C_{60,60}=137.5\ $GHz$\,\mu$m$^6$. Note that Penning ionization~\cite{MX_WEIDEMULLERPAIRPENNING07,MX_WEIDEMULLERANTIBLOC10} is then negligible (the free ions collected at the onset of the field ionization ramp are $\simeq 1$\% of the Rydberg signal). 

The spatial distribution of the Rydberg atoms in the sample and thus their interaction energy can be controlled by adjusting the detuning $\Delta$ of the blue laser with respect to the resonance frequency of non-interacting atoms~\cite{MX_WEIDEMULLERPAIRPENNING07,MX_EVERSRYDAGGREGATE13,MX_PILLETDISSRYD14,MX_OTTSUPERATOM15}. In a simple incoherent excitation model at resonance ($\Delta =0$), dipole blockade precludes the excitation of more than one Rydberg atom  inside a sphere with radius $R_b=(C_{60,60}/\gamma)^{1/6}\simeq 8\ \mu$m. For a blue detuning ($\Delta> 0$), after the non-resonant excitation of a first `seed' Rydberg atom, the excitation of a second one is resonantly `facilitated' on a spherical surface of radius $R_f= (C_{60,60}/\Delta)^{1/6}$ ~\cite{MX_EVERSRYDAGGREGATE13}. As additional atoms get excited, the facilitation surface increases. An avalanche process forms a cluster around the seed atom. If $\Delta>\gamma$, $R_f<R_b$ and the mean interatomic distance is reduced. The interaction energy per atom thus increases with $\Delta$. 

\begin{figure}[tbm]
\centering
\includegraphics[width=0.48\textwidth]{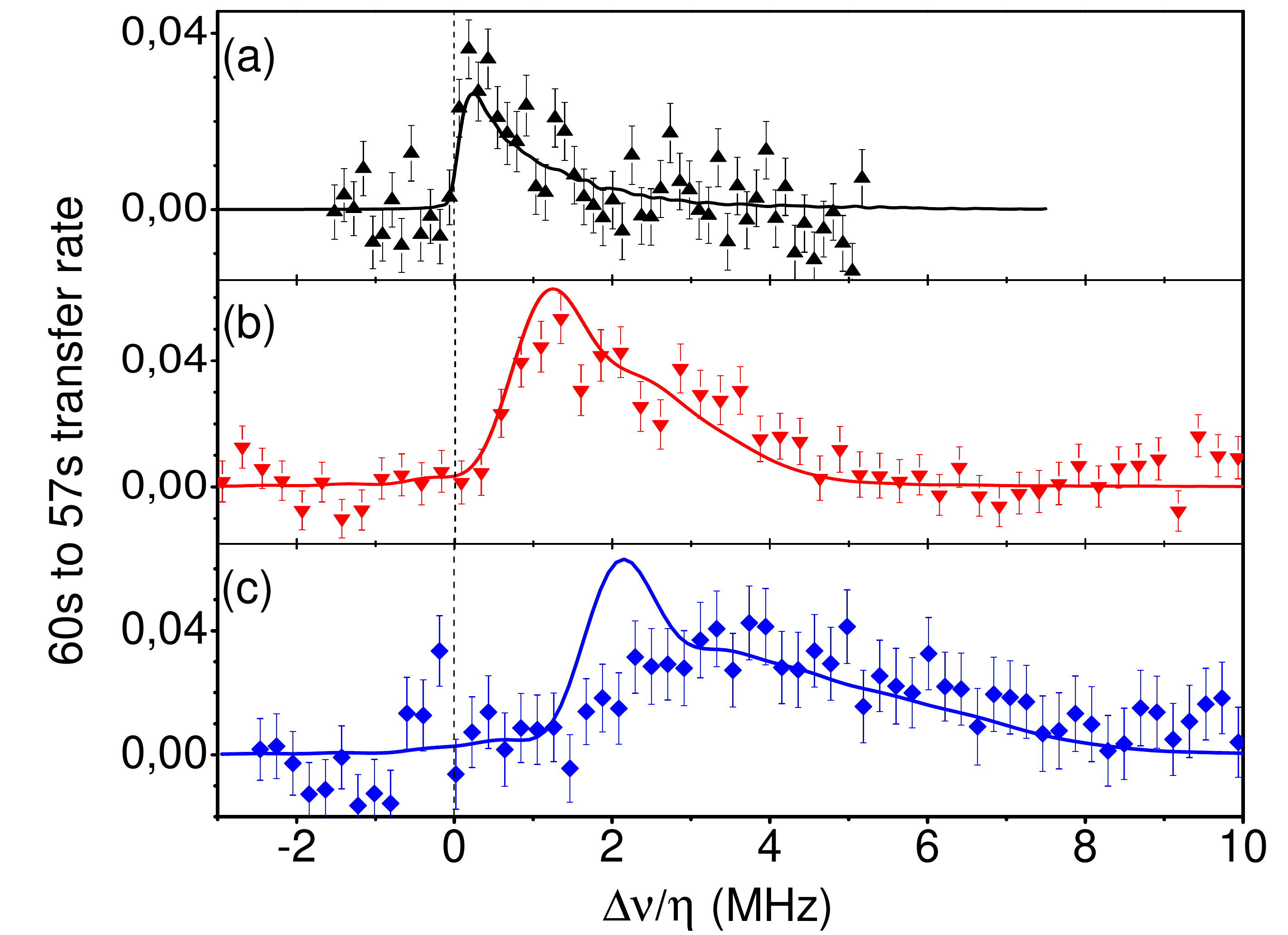}
\caption{(color online). Microwave spectra of the $60S-57S$ two-photon transition for (a) $\Delta= 0$ (b) $\Delta= 1$~MHz, (c)  $\Delta= 2$~MHz. The dots are experimental with statistical standard deviation error bars and the solid line results from a simple model. The origin of the frequency axis (thin vertical line) corresponds to the resonance position in a low-density cloud. } 
\label{MWspectra}
\end{figure}

We probe the interaction energy distribution by microwave spectroscopy. We choose  a 60S$\rightarrow$nS two-photon transition, with a narrow (few kHz width) spectral line at frequency $\nu_0/2$ for non-interacting atoms~\cite{ENS_CHIPSPECTRO14}. Atomic interactions shift and broaden this line. Let us consider first a pair of atoms at a distance $R$. A weak microwave couples the initial $|60S,60S\rangle$ state only to $|60S,nS\rangle$ and $|nS,60S\rangle$ states. The latter are resonantly coupled by the vdW interaction. The interaction matrix involves diagonal, $hC_{60,n}/R^6$, and off-diagonal exchange terms, $hA_{60,n}/R^6$. The eigenstates are symmetric and antisymmetric combinations of the bare states, with energies separated by $2h A_{60,n}/R^6$. We thus generally get two excitation lines for a single pair. The situation is much simpler if we choose $n=57$ ($\nu_0=116.457$~GHz). The exchange term, $A_{60,57}= -0.33\ $GHz\,$\mu$m$^6$, is then much smaller than $C_{60,57}= -43.3\ $GHz$\,\mu$m$^6$. We get a single excitation line, at frequency $[\nu_0+\Delta \nu(R)]/2$, with  $\Delta \nu(R)= (C_{60,60}-C_{60,57})/R^6 = \eta\,C_{60,60}/R^6$, where $\eta\simeq 1.316$.

For $N$ interacting atoms, an immediate generalization leads to a frequency shift $\Delta \nu^{(i)}/2$ for atom number $i$ (assumed to be far away from all other atoms in state 57S), with
\begin{eqnarray}
\Delta \nu^{(i)} &=& (C_{60,60}-C_{60,57})\sum_{1\le j\le N\atop i\ne j}1/R_{ij}^6 \nonumber\\ &=& \eta \,\Delta E^{(i)}/h \ ,
\end{eqnarray}
where $\Delta E^{(i)}$ is the vdW energy shift of atom $i$ in the $60S$ state due to its interaction with all other Rydberg levels. 

  
We first probe the spectrum with a 1~$\mu$s microwave pulse at an adjustable frequency, $(\nu_0+\Delta\nu)/2$, applied immediately after laser excitation, so that atomic motion plays no role. We use three laser detunings, $\Delta=0,\ 1$ and 2~MHz, resulting in about 80, 60 and 40 detected Rydberg atoms for each laser pulse respectively. Less than $\approx 3$ atoms are transferred into the 57S state. Direct interactions between these 57S atoms can be neglected. Since the vdW  57S--60S interaction is attractive, some 57S atoms (less than 50\%) undergo Penning ionization~\cite{MX_WEIDEMULLERPAIRPENNING07}. This process, taking place after the microwave pulse, reduces the measured transfer rate but does not alter the spectrum shape.

Figure \ref{MWspectra} presents, for the three $\Delta$ values, the fraction of atoms detected in $57S$ as a function of the scaled microwave frequency, $\Delta\nu/\eta$. The points are experimental and the solid lines result from a Monte Carlo simulation assuming successive incoherent Rydberg excitations from the ground state. At each iteration step, we select randomly a ground state atom according to the trap geometry. We randomly choose whether this atom is excited or not, taking into account the vdW energy shift due to previously excited atoms, the full laser linewidth and intensity profile. The number of iterations is chosen to reproduce the observed average Rydberg atom number. The final microwave spectrum is a convolution of the interaction energy distribution with the microwave pulse linewidth. We average 50 to 200 Monte Carlo realizations and adjust the vertical scale so that the area of the spectrum fits the experimental one. This procedure provides us with the solid lines in Fig. \ref{MWspectra}. They are in fair agreement with the experiment. 

The $\Delta=0$ spectrum maximum  [Fig. \ref{MWspectra}(a)] is frequency-shifted by about the laser linewidth $\gamma$. Hence, the interatomic distances are close to the dipole blockade radius $R_b$. The high frequency tail is due to atoms in the cloud bulk, with several neighbors at a distance close to $R_b$. We infer from the simulation that the size of the Rydberg atom ensemble is three times larger than that of the ground state cloud. This large broadening results from dipole blockade. Since atoms cannot be excited at short distances, laser excitation favors the tail of the thermal cloud Gaussian distribution. This precludes the observation of a clear Rydberg atom number saturation with a thermal cloud.

For  $\Delta= 1$ and 2 MHz [Fig. \ref{MWspectra}(b) and (c)], the  total Rydberg atom number is reduced due to the non-resonant excitation. Nevertheless, the interaction energy increases with $\Delta$, corresponding to an increasing peak Rydberg density. The average shift is of the order of $2\Delta$, as expected from an energy conservation argument stating that $1/2 \sum_{i=1}^{N} E^{(i)}\simeq N.h\Delta$. The agreement with the Monte Carlo simulation is good for $\Delta= 1$~MHz. For $\Delta=2$~MHz, the simulation predicts a bimodal structure with a narrow component centered on $\Delta$. This peak is due to atoms excited on the outer `facilitation' surface, with an initial interaction energy $h\Delta$ unmodified since there is no further excitation in their vicinity. Our simple model, which does not take into account coherent pair excitation processes, overestimates this contribution.     

\begin{figure}[tbm]
\centering
\includegraphics[width=0.48\textwidth]{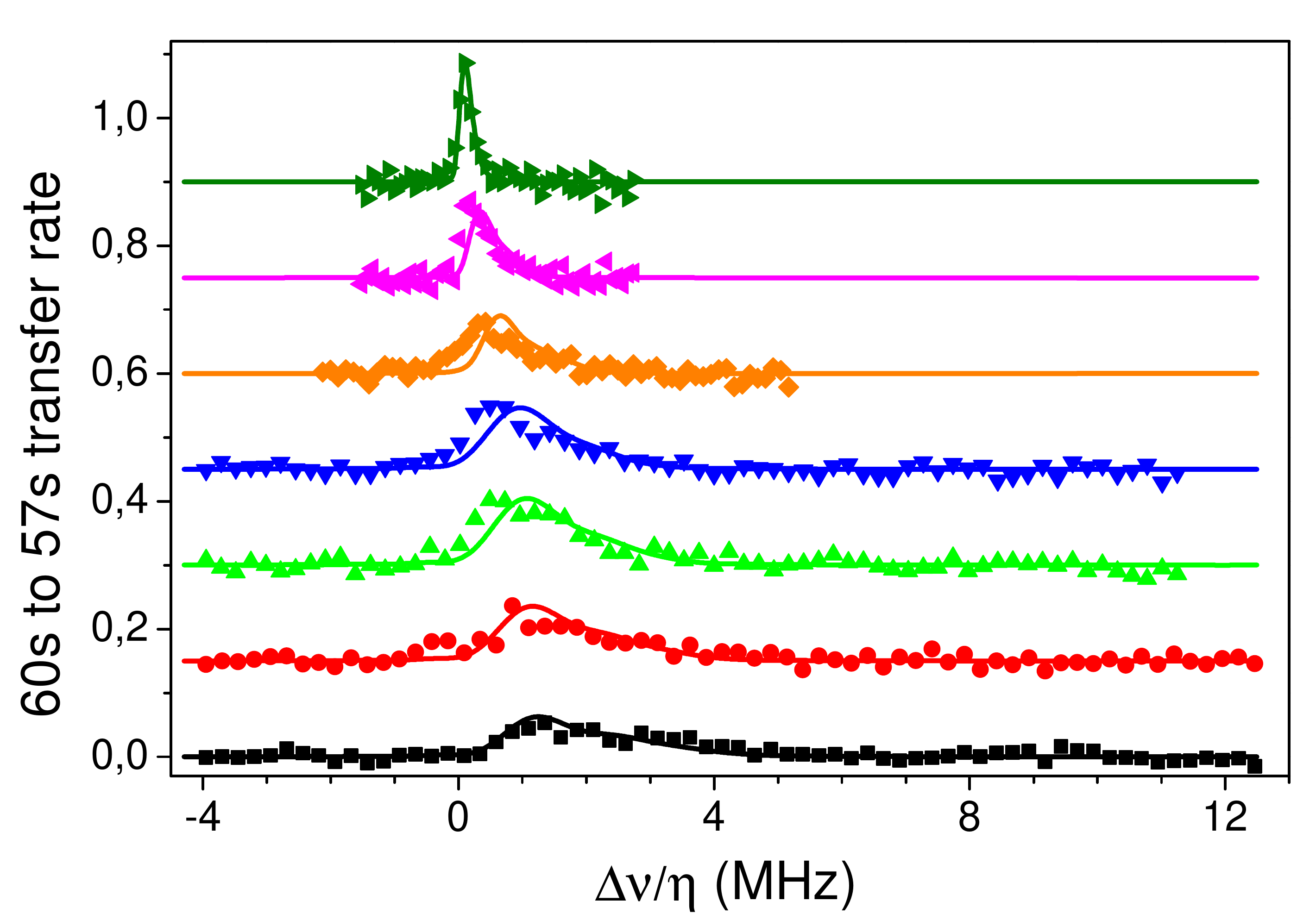}
\caption{(color online) Rydberg atoms interaction energy spectrum as a function of the delay $\tau$ between atomic preparation and microwave probe (0.5, 10, 15, 20, 30, 50, 100~$\mu$s from bottom to top). The points are experimental (error bars not shown for clarity) and the lines result from the model (see text). The different spectra are shifted vertically by a fixed offset (0.15) for the sake of clarity.} 
\label{expansion}
\end{figure}

We then investigate the Rydberg gas expansion due to the repulsive vdW forces between 60S atoms. Figure \ref{expansion} presents (points) measurements of the interaction energy spectra at different delays between the laser and the microwave pulses for $\Delta=1\,$MHz. For this detuning, the internal potential energy per atom is of the order of $50\,\mu$K, two orders of magnitude above the initial temperature. As $\tau$ increases, the spectrum gets narrower and converges to that for a dilute cloud with negligible interactions. The theoretical lines in Fig. \ref{expansion}  result from the Monte Carlo excitation model, followed by a direct integration of the Newtonian equations of motion. The calculation takes into account the finite lifetime ($210\,\mu$s~\cite{ENS_CHIPSPECTRO14}) of the 60S state, even though the atomic decay contribution to the interaction energy reduction is small (at most 30\%). The agreement with the experimental data is good, reinforcing our confidence in the excitation model.

\begin{figure}[h]
\centering
\includegraphics[width=0.5\textwidth]{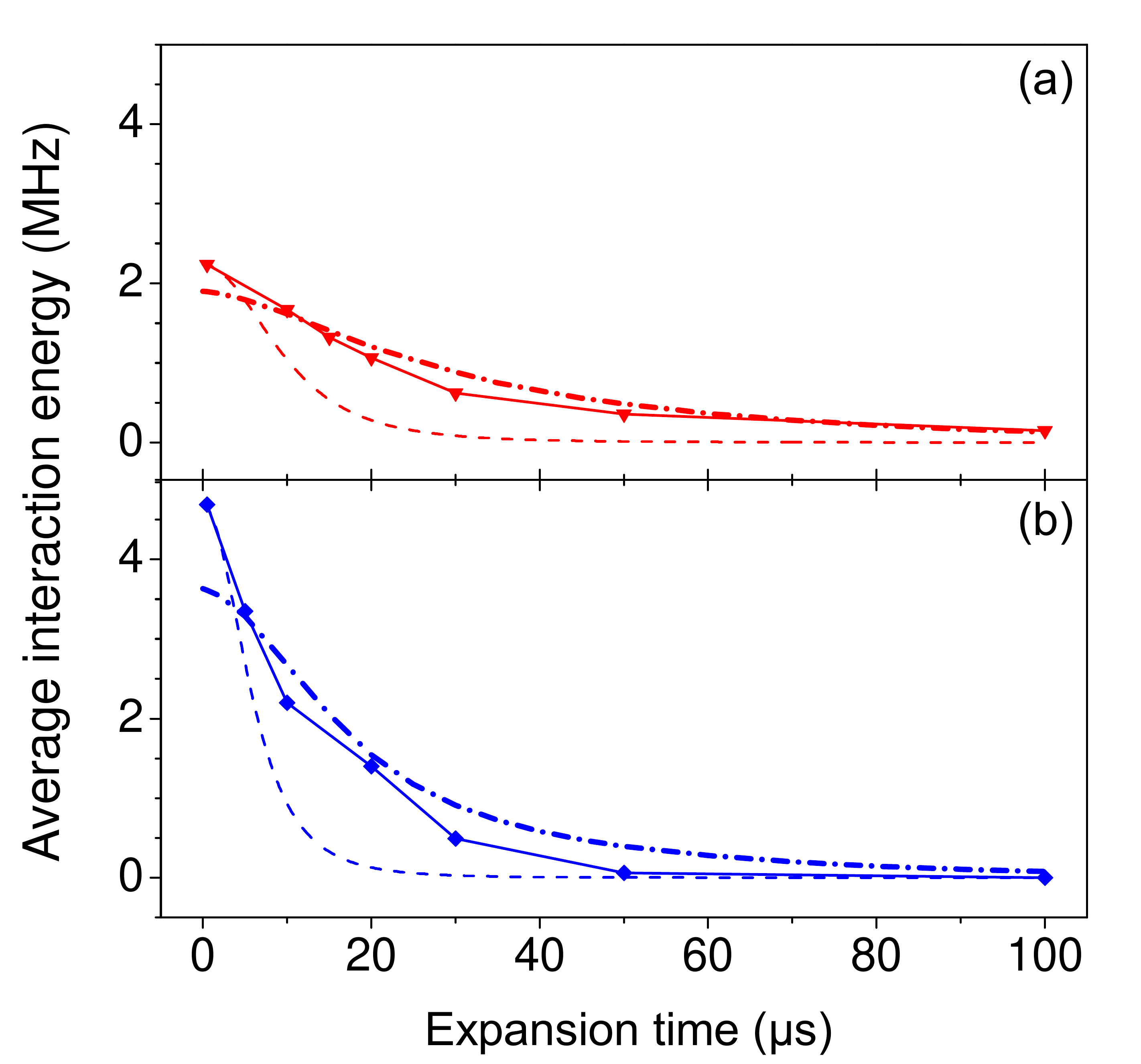}
\caption{(color online) Variation of the average interaction energy as a function of time for (a) $\Delta= 1$~MHz and (b) $\Delta= 2$~MHz. The points (connected by a solid line for visual convenience) are experimental. The thick dot-dashed line results from our model. The dashed line corresponds to a single Rydberg atom pair expansion model.} 
\label{av_energy}
\end{figure}

We plot in figure \ref{av_energy} the time evolution of the average interaction energy for $\Delta= 1$ and $2\,$ MHz (points and connecting solid lines). We observe, as expected, a decay of the vdW potential energy at the benefit of atomic kinetic energy, faster when the initial energy is larger. The dot-dashed lines present the results of the Monte Carlo model, in fair agreement with the data. Since the model slightly underestimates the initial energy, particularly for $\Delta=2$~MHz, it predicts a slightly slower expansion. The experimental expansion is much slower than that of a single Rydberg atom pair with the same average initial energy (dashed lines in figure \ref{av_energy}). The expansion proceeds in an interesting hydrodynamic regime, requiring more detailed studies.

We use microwave spectroscopy for a direct measurement of the vdW interaction energy distribution in a cold atom sample containing up to 80 interacting Rydberg atoms. The spectra provide direct evidence of the dipole blockade regime. Time-resolved measurements give access to the dynamical evolution of the Rydberg atom cloud, whose fast expansion sets limits to the frozen gas approximation. 

This new diagnostic method leads to a direct observation of self-organization and dynamical phase transitions in Rydberg atom-based quantum simulators~\cite{MX_ZOLLERRYDBERGSIMUL12}. The strong confinement possible with a superconducting atom chip opens the way to the realization of unidimensional atomic clouds, with long Rydberg coherence times~\cite{ENS_CHIPSPECTRO14} and, hence, to quantum simulations of interacting spin chains~\cite{MX_LESANOVSKYSPINCHAIN12,MX_WEIDEMULLERRYDBERG13,MX_FLEISCHAUERRATE13,MX_FLEISCHHAUERANTIFERRO14}. In particular, the strong exchange vdW interaction between 60S and 61S levels could be used to simulate quantum transport processes~\cite{MX_HORODECKISPINCHAIN14}.

\begin{acknowledgements}

This research has been supported by the EU Marie Curie Action CCQED, Project 264666, by the EU ICT Project SIQS Number 600645 and by the DECLIC ERC project. 
\end{acknowledgements}


\end{document}